\documentclass{basi}
\usepackage{epsfig}
\usepackage[bookmarks=false]{hyperref}

\begin{document}
\title[Estimation of mass and cosmological constant]
      {Estimation of mass and cosmological constant of nearby spiral galaxies using galaxy rotation curve}
\author[B. Aryal et al.]%
       {B. Aryal,$^{\rm 1,2}$\thanks{e-mail:binil.aryal@uibk.ac.at}, R. Pandey,$^{\rm 2}$ N. Baral,$^{\rm 2}$ U. Khanal$^{\rm 2}$ and W. Saurer$^{\rm 1}$\\
        $^{\rm 1}$Institute of Astro- and Particle Physics, Innsbruck University, Technikerstrasse 25\\A-6020 Innsbruck, Austria\\
        $^{\rm 2}$Central Department of Physics, Tribhuvan University,
Kirtipur, Kathmandu, Nepal\\}
\date{Received 2013 January 22; accepted 2013 ..........}

\maketitle

\label{firstpage}
\begin{abstract}
An expression for rotational velocity of a test particle around
the central mass in the invariant plane is derived. For this, a
line element of Schwarzschild de-Sitter space-time is used to
study the effect of cosmological constant ($\Lambda$) on the
motion of both massive and massless particles. Using rotation
curve data of 15 nearby spiral and barred spiral galaxies, we
estimated the mass and $\Lambda$ of the galaxy. The mass of the
galaxies are found to lie in the range 0.13-7.60 $\times$
10$^{40}$ kg. The cosmological constant ($\Lambda$) is found to be
negative ($-$0.03 to $-$0.10 $\times$ 10$^{-40}$ km$^{-2}$),
suggesting the importance of anti-de Sitter space in the local
bubble. Possible explanation of the results will be discussed.
\end{abstract}


\section{Introduction}
\label{sec:intro} The cosmological constant ($\Lambda$) is a
parameter describing the energy density of the vacuum (empty
space). A negative $\Lambda$ adds to the attractive gravity of
matter whereas a positive $\Lambda$ resists the attractive gravity
of matter due to its negative pressure. Two supernova cosmology
project (Perlmutter et al. 1999; Schmidt et al. 2000) converge on
a common result: cosmos will expand forever at an accelerating
rate, pushed on by dark energy ($\Lambda$). According to this, the
Universe is discovered to have a flat spatial geometry ($k$ = 0),
a positive cosmological constant (dark energy) $\Omega_{\Lambda}$
= 0.7, and matter density (of all kinds) $\Omega_{M}$ = 0.7
(Perlmutter 2003). The data from the Boomerang experiment provide
a firm evidence for the case of flat Universe $\Omega_{k}$ $\sim$
0 (Perlmutter et al. 1999). Overduin et al. (2000) discussed some
of the theoretical justifications for a non zero $\Lambda$.
Kraniotis \& Whitehouse (2000) interpreted $\Lambda$ as the
non-gravitational contribution to galactic velocity rotation
curves and estimated negative $\Lambda$ for 6 spiral galaxies,
suggesting de-accelerating universe. The implications of this
result will be discussed later.

In the present work, we discuss an expression for rotational
velocity of a test particle in a circular motion around the
central mass in an asymptotically de Sitter space time (Pokheral
1994). This expression is used to estimate the value of the
$\Lambda$ from the rotation curve of various spiral galaxies. We
estimated the value of $\Lambda$ for 15 spiral galaxies that have
radial velocity less than 2\,000 km s$^{-1}$. This paper is organized as follows.
A relation between rotational velocity and the cosmological constant is derived in Sect. 2.
We describe the database and the calculation procedure in Sects. 3
and 4. In Sect. 5, we present our result. Finally, a discussion of
results and the conclusions are summarized in Sects. 6 and 7.

\section{Theory}

The metric which describes the geometry of the Schwarzschild
de-Sitter space-time is
\begin{equation}
\begin{array}{l}
ds^{2}$=$(1-\frac{2M}{r}-\frac{\Lambda
r^{2}}{3})dt^{2}-\frac{dr^{2}}{1-\frac{2M}{r}-\frac{\Lambda
r^{2}}{3}}-r^{2}(d\theta^{2}+\sin^{2}\theta d\varphi)
\end{array}
\end{equation}
Here we utilize this line element in order to study the effect of
the cosmological constant $\Lambda$ on the motion of both massive
and massless particles in the Universe. For this space-time, the
Lagrangian is
\begin{equation}
\begin{array}{l}
\mathcal L=\frac{1}{2}[(1-\frac{2M}{r}-\frac{\Lambda
r^{2}}{3})\dot{t}^{2}-\frac{\dot{r^{2}}}{1-\frac{2M}{r}-\frac{\Lambda
r^{2}}{3}}-r^{2}\dot{\theta}^{2}-r^{2}\sin^{2}\theta\dot{\varphi}^{2}]
\end{array}
\end{equation}
where dot represents differentiation with respect to the affine
parameter $\tau$. The canonical momenta are
\begin{equation}
\begin{array}{l}
\wp_{t}=\frac{\partial\mathcal
L}{\partial\dot{t}}=(1-\frac{2M}{r}-\frac{\Lambda
r^{2}}{3})\dot{t}
\end{array}
\end{equation}
\begin{equation}
\begin{array}{l}
\wp_{\theta}=-\frac{\partial\mathcal
L}{\partial\dot{\theta}}=r^{2}\dot{\theta}
\end{array}
\end{equation}
\begin{equation}
\wp_{r}=-\frac{\partial\mathcal
L}{\partial\dot{r}}=(1-\frac{2M}{r}-\frac{\Lambda
r^{2}}{3})^{-1}\dot{r}
\end{equation}
and
\begin{equation}
\wp_{\varphi}=-\frac{\partial\mathcal
L}{\partial\dot{\varphi}}=r^{2}\sin^{2}\theta\dot{\varphi}
\end{equation}
The resulting Hamiltonian is
\begin{equation}
\mathcal
H=\wp_{t}-(\wp_{r}\dot{r}+\wp_{\theta}\dot{\theta}+\wp_{\varphi}\dot{\varphi})-\mathcal
L=\mathcal L
\end{equation}
The equality of Hamiltonian and Lagrangian signifies that there is
no `potential' in the problem. Thus, the energy is derived solely
from `kinetic energy'.

Now we rescale the affine parameter $\tau$ in such away that the
2$\mathcal L$ has the value +1 for time-like geodesics and zero
for null geodesics. In this case the integrals of motion in the
invariant plane $\theta=\frac{\pi}{2}$ are
\begin{equation}
\wp_{t}=(1-\frac{2M}{r}-\frac{\Lambda r^{2}}{3})\dot{t}= constant
= E
\end{equation}
and \begin{equation}
\wp_{\varphi}=r^{2}\dot{\varphi}= constant =
L
\end{equation}
where $E$ and $L$ are constants associated with the energy and
angular momentum of the particle (about an axis normal to the
invariant plane) respectively.

Using $\dot{t}$ and $\dot{\varphi}$ (equations 8 and 9), the
constancy of the Lagrangian gives
\begin{equation}
\begin{array}{l}
\frac{E^{2}}{1-\frac{2M}{r}-\frac{\Lambda
r^{2}}{3}}-\frac{\dot{r}}{1-\frac{2M}{r}-\frac{\Lambda
r^{2}}{3}}-\frac{L^{2}}{r^{2}}=2\mathcal L
\\
\\\,\,\,\,\,\,\,\,\,\,\,\,\,\,\,\,\,\,\,\,\,\,\,\,\,\,\,\,\,\,\,
\,\,\,\,\,\,\,\,\,\,\,\,\,\,\,\,\,\,\,\,\,\,\,\,\,\,\,\,\,\,\,
\,\,\,\,\,\,\,\,\,\,\,\,\,\,\,\,\, = +1\,\,\, $or$\,\,\, 0
\end{array}
\end{equation}
depending upon whether we are considering time-like or null
geodesics. Defining further a parameter $\Omega$, where,
\begin{equation}
\Omega=\frac{d\varphi}{dt}=\frac{\dot{\varphi}}{\dot{t}}=\frac{L}{E}(\frac{1-\frac{2M}{r}-\frac{\Lambda
r^{2}}{3}}{r^{2}}),
\end{equation}
The rotational velocity of a test particle around the central mass
in the invariant plane is given by (Chandrashekhar 1983),
\begin{equation}
v_{\varphi}=\frac{r}{\sqrt{1-\frac{2M}{r}}-\frac{\Lambda
r^{2}}{3}}\Omega
\end{equation}
For time-like geodesics, equations (10) and (11) can be rewritten
in the form
\begin{equation}
(\frac{dr}{d\tau})-(1-\frac{2M}{r}-\frac{\Lambda
r^{2}}{3})(1+\frac{L^{2}}{r^{2}})+E^{2} = f(r)
\end{equation}
and
\begin{equation}
\frac{d\varphi}{d\tau}=\frac{L}{r^{2}}
\end{equation}
Dividing equation (13) by the $(\frac{d\varphi}{dt})^2$ from
equation (14). We can get the equation for $r$ as a function of
$\varphi$:
\begin{equation}
(\frac{dr}{d\varphi})^{2}=\frac{\Lambda
r^{6}}{3L^{2}}+(E^{2}+\frac{\Lambda
L^{2}}{3}-1)\frac{r^{4}}{L^{2}}+\frac{2Mr^{3}}{L^{2}}-r^{2}+2Mr
\end{equation}
Now letting $u=\frac{1}{r}$, we obtain the basic equation of the
problem
\begin{equation}
\begin{array}{l}
(\frac{du}{d\varphi})^{2}=2Mu^{3}-u^{2}+\frac{2Mu}{L^{2}}+\frac{\Lambda}{3L^{2}u^{2}}-(\frac{1-E^{2}}{L^{2}}-\frac{\Lambda}{3})
\\\,\,\,\,\,\,\,\,\,\,\,\,\,\,\,\,=f(u)
\end{array}
\end{equation}
Since $f(u)$ involves a polynomial of fifth degree (i.e., quantic
polynomial). The exact solution to the problem cannot be
formulated in terms of elliptic integrals (Gradshteyn \& Ryzhik
1997, Goldstein 1980).

For circular orbits, $f(r)$ in equation (13) should have double
root. The critical values of $L$ and $E$ required for such double
roots can be found, by setting $f(r)$ and $f'(r)$ equal to zero.
Accordingly,
\begin{equation}
f(r)=E^{2}-(1-\frac{2M}{r}-\frac{\Lambda
r^{2}}{3})(1+\frac{L^{2}}{r^{2}})=0
\end{equation}
and
\begin{equation}
\begin{array}{l}
f'(r)=\frac{df}{dr}\\
\\
\,\,\,\,\,\,\,\,\,\,\,\,\,\,\,=-(1-\frac{2M}{r}-\frac{\Lambda
r^{2}}{3})(\frac{-2L^{2}}{r^{3}})-(1+\frac{L^{2}}{r^{2}})(\frac{2M}{r^{2}}-\frac{2\Lambda
r}{3})\\
\\\,\,\,\,\,\,\,\,\,\,\,\,\,\,\,=0
\end{array}
\end{equation}
Eliminating the factor $(1+\frac{L^{2}}{r^{2}})$, we get
\begin{equation}
 \frac{L}{E}=\frac{r\sqrt{\frac{M}{r}-\frac{\Lambda r^{2}}{3}}}
 {1-\frac{2M}{r}-\frac{\Lambda r^{2}}{3}}
\end{equation}
Equation (11) reduces to
\begin{equation}
    \Omega=\frac{\sqrt{\frac{M}{r}-\frac{\Lambda r^{2}}{3}}}{r}
\end{equation}
By substituting equation (20) in equation (12), we find
\begin{equation}
 v_{\phi}=\sqrt{\frac{\frac{M}{r}-\frac{\Lambda r^{2}}{3}}
 {1-\frac{2M}{r}-\frac{\Lambda r^{2}}{3}}}
\end{equation}
This equation represents the rotational velocity of test particle
around the central mass in the invariant plane. This expression
can be used to determine the mass and the cosmological constant
($\Lambda$) of galaxies using rotation curve data.

\section{Database}

A galaxy needs to fulfill the following selection criteria in
order to be selected: (1) galaxy rotation curve should be given,
(2) radial velocity $<$ 2\,000 km s$^{-1}$ and, (3) morphology
should be known. In addition to this, the diameters, position
angle and magnitude of the galaxies were added.
Table 1 lists the database. The rotation curve data of these
galaxies are taken from Sofue (2007). In the database, we have
added a blue-shifted galaxy (NGC 0224 / RV = --300 km s$^{-1}$) in
order to observe the local effect. There are 11 spirals and 4
barred spirals. The late-type galaxies (Sc) dominate (6, 40\%) the
database. The radial velocity of the red-shifted galaxies lie in
the range 131 km s$^{-1}$ (NGC 2403) to 1\,636 km s$^{-1}$ (NGC
1365).

\begin{table*}
\centering \caption[]{The database of 15 galaxies (Sanders 1996;
Sofue et al. 1999, Barnaby \& Thronson 1992, 1994; Sancisi \& van
Albada 1987; Kent 1985). The first three columns list the NGC name
and positions of galaxies. The next two columns give the major
($a$) and minor ($b$) diameters (in arcmin). The sixth, seventh
and eight columns list the radial velocity ($RV$, in km s$^{-1}$),
rotational velocity ($V_{rot}$, in km s$^{-1}$) and distance ($D$,
in Mpc). The last two columns give the position angle ($PA$, in
degree) and the morphology ($T$) of galaxies.}
$$
\begin{array}{p{0.10\linewidth}lcccccccccccc}
\hline \noalign{\smallskip}
   NGC   & $RA (J2000)$   & $Dec (J2000)$   &  a   & b    &  RV     & V_{rot} &  D  & PA   &  $T$ \\
  \hline
  0224   & 00^{\rm h}40^{\rm m}00.09^{\rm s}  & +40^{\circ}59'42.8''   & 190.0  & 60.0   & -300    & 250   & 0.69    & 40   & $Sb$ \\
  0660   & 01^{\rm h}43^{\rm m}02.40^{\rm s}  & +13^{\circ}38'42.2''   & 08.3   & 03.2   & 850     & 145   & 13.00   & 45   & $Sc$  \\
  0891   & 02^{\rm h}22^{\rm m}33.41^{\rm s}  & +42^{\circ}20'56.9''   & 13.5   & 02.5   & 528     & 162   & 8.90    & 19   & $Sb$ \\
  1097   & 02^{\rm h}46^{\rm m}19.05^{\rm s}  & -30^{\circ}16'29.6''   & 09.3   & 06.3   & 1271    & 283   & 16.00   & 135  & $SBb$ \\
  1365   & 03^{\rm h}33^{\rm m}36.37^{\rm s}  & -36^{\circ}08'25.4''   & 11.2   & 06.2   & 1636    & 235   & 15.60   & 222  & $SBb$ \\
  2403   & 07^{\rm h}36^{\rm m}51.40^{\rm s}  & +65^{\circ}36'09.2''   & 21.9   & 12.3   & 131     & 137   & 3.30    & 125  & $Sc$ \\
  2903   & 09^{\rm h}32^{\rm m}10.11^{\rm s}  & +21^{\circ}30'03.0''   & 12.6   & 06.0   & 556     & 252   & 6.10    & 21   & $Sc$ \\
  3198   & 10^{\rm h}19^{\rm m}54.92^{\rm s}  & +45^{\circ}32'59.0''   & 10.0   & 03.8   & 663     & 163   & 9.10    & 215  & $SBc$ \\
  4258   & 12^{\rm h}18^{\rm m}57.50^{\rm s}  & +47^{\circ}18'14.3''   & 18.6   & 07.2   & 448     & 210   & 6.60    & 150  & $Sbc$ \\
  4321   & 12^{\rm h}22^{\rm m}54.90^{\rm s}  & +15^{\circ}49'20.6''   & 07.4   & 06.3   & 1571    & 271   & 15.00   & 146  & $Sc$ \\
  4565   & 12^{\rm h}30^{\rm m}20.78^{\rm s}  & +25^{\circ}59'15.6''   & 15.9   & 01.9   & 1230    & 252   & 10.20   & 137  & $Sb$ \\
  5033   & 13^{\rm h}13^{\rm m}27.53^{\rm s}  & +36^{\circ}35'38.1''   & 10.7   & 05.0   & 875     & 183   & 14.00   & 179  & $Sc$ \\
  5055   & 13^{\rm h}15^{\rm m}49.33^{\rm s}  & +42^{\circ}01'45.4''   & 12.6   & 07.2   & 504     & 182   & 8.00    & 103  & $Sbc$ \\
  5236   & 13^{\rm h}37^{\rm m}00.92^{\rm s}  & -29^{\circ}51'56.7''   & 12.9   & 11.5   & 513     & 178   & 8.90    & 45   & $SBc$ \\
  5907   & 15^{\rm h}15^{\rm m}53.77^{\rm s}  & +56^{\circ}19'43.6''   & 12.8   & 01.4   & 667     & 245   & 11.60   & 156  & $Sc$ \\
 \hline
\end{array}
$$
\end{table*}

\begin{figure*}
\vspace{0.0cm} \centering
\includegraphics[height=2.45cm]{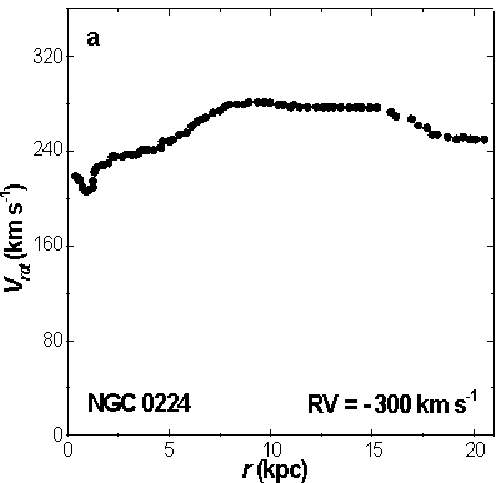}
\includegraphics[height=2.45cm]{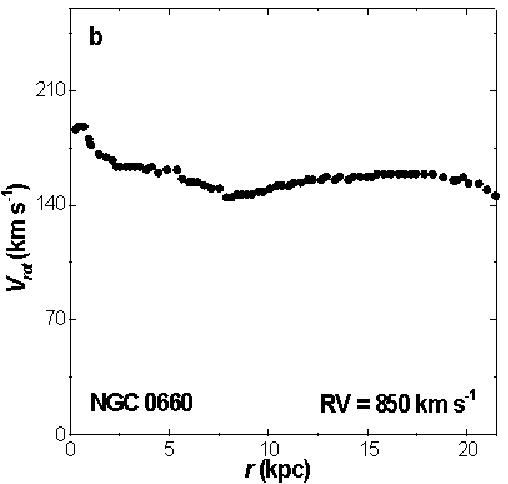}
\includegraphics[height=2.45cm]{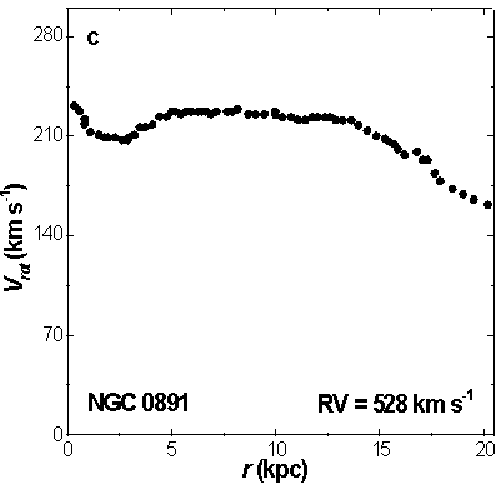}
\includegraphics[height=2.45cm]{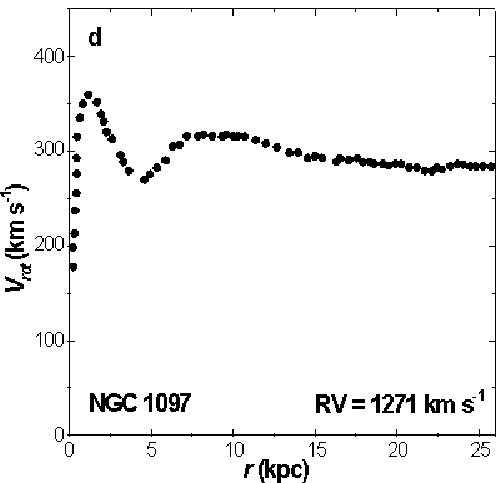}
\includegraphics[height=2.45cm]{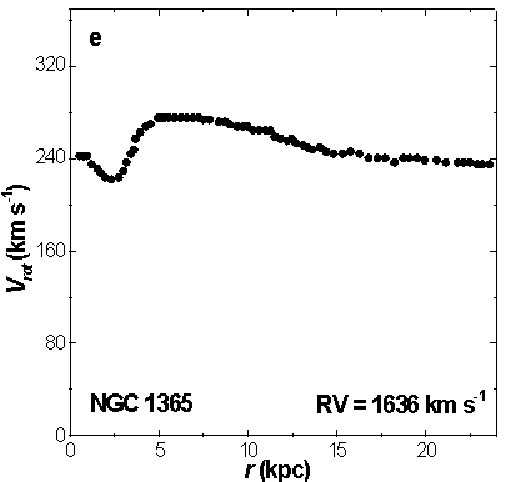}
\includegraphics[height=2.45cm]{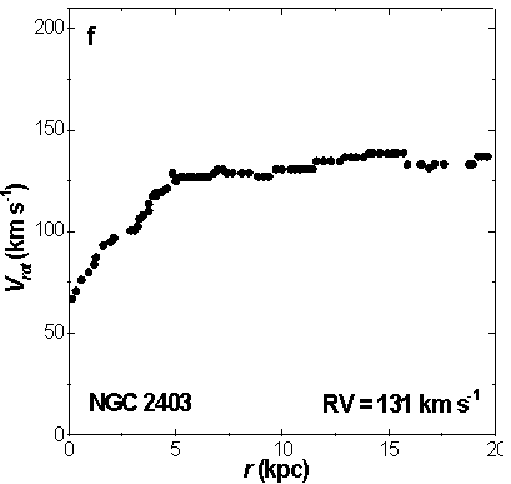}
\includegraphics[height=2.45cm]{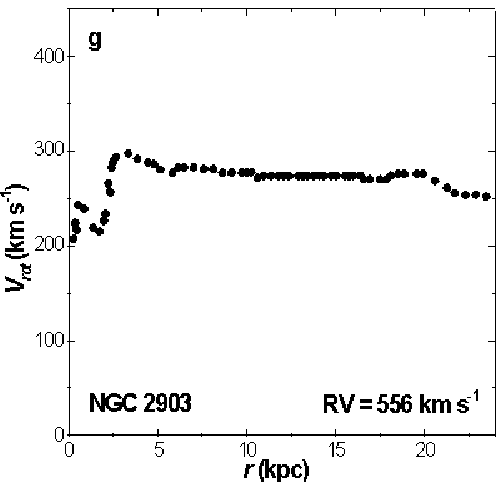}
\includegraphics[height=2.45cm]{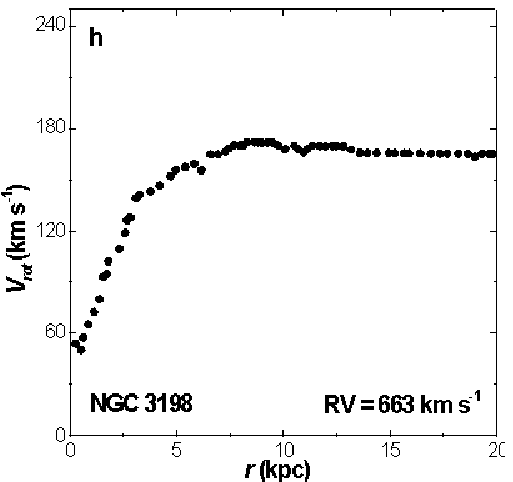}
\includegraphics[height=2.45cm]{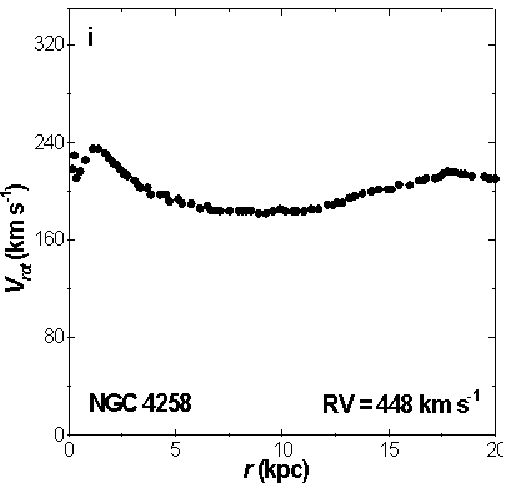}
\includegraphics[height=2.45cm]{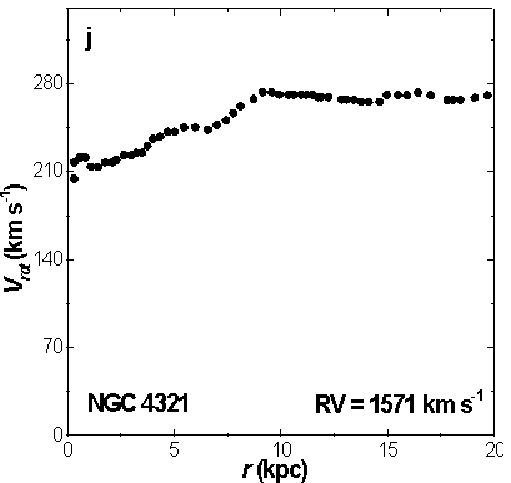}
\includegraphics[height=2.45cm]{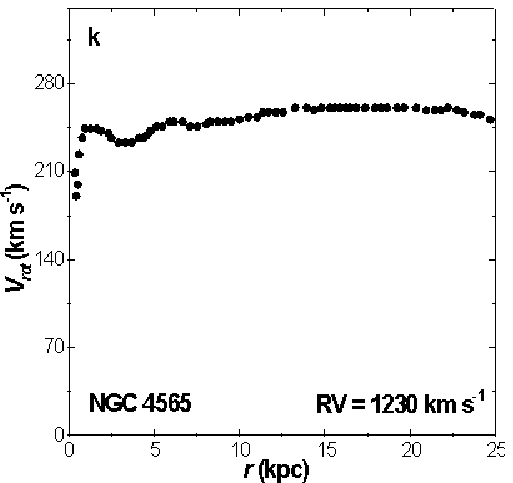}
\includegraphics[height=2.45cm]{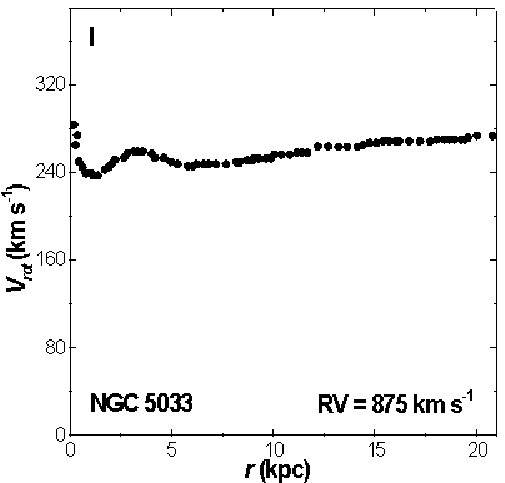}
\includegraphics[height=2.45cm]{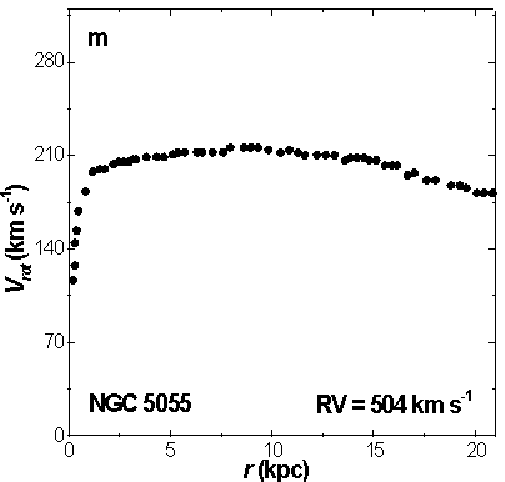}
\includegraphics[height=2.45cm]{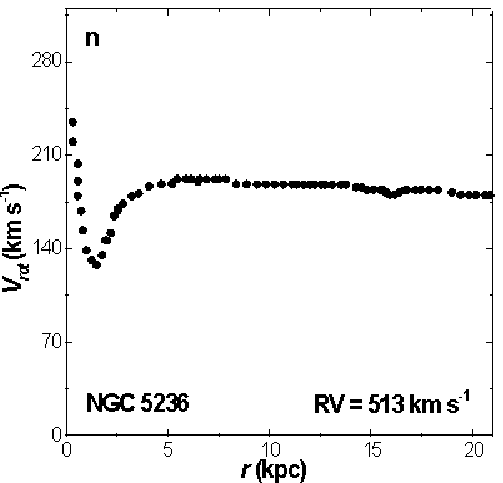}
\includegraphics[height=2.45cm]{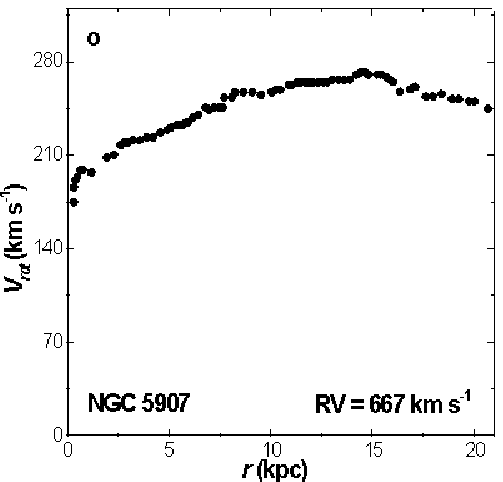}
\caption{Galaxy rotation curves of individual galaxies (Sofue
2007). The NGC name and the $RV$ of the galaxies are given.}
\end{figure*}

The rotation curves of these galaxies comprise steep central rise,
bulge component, broad maximum by the disk and the halo component
(Fig. 1). It can be seen that the rotational velocities in many
well resolved galaxies do not declined to zero at the nucleus.
This hints that the mass density increases rapidly towards the
nucleus than the expected from exponential or de Vaucouleurs laws.
The widely adopted zero velocity at the center is due to the
linkage (merely a drawing) between positive and negative sides of
the nucleus. However, we focus our attention in the flat region of
the rotation curve.

We use the value of $r$ (in kpc) and $v$$_{rot}$ (in km s$^{-1}$)
from the given rotation curves in order to estimate the value of
the mass ($M$) and cosmological constant ($\Lambda$) of galaxies.

\section{Estimation of $\Lambda$ and $M$}

The rotational velocity of a test particle around the central mass
in the invariant plane is given by,
\begin{equation}
\begin{array}{l}
\frac{v_{\phi}}{c}=[\frac{\frac{M}{r}-\frac{\Lambda
r^{2}}{3}}{1-\frac{2M}{r}-\frac{\Lambda r^{2}}{3}}]^{\frac{1}{2}}
\end{array}
\end{equation}
Then,
\begin{equation}
\begin{array}{l}
\frac{v^{2}_{\phi}}{c^2}(1-\frac{2M}{r}-\frac{\Lambda r^{2}}{3})=
(\frac{M}{r}- \frac{\Lambda r^{2}}{3})
\end{array}
\end{equation}
Taking summation,
\begin{equation}
\begin{array}{l}
\Sigma
v^{2}_{\phi}(\frac{1}{c^2})-2M\Sigma\frac{v^{2}_{\phi}}{r}(\frac{1}{c^2})-
\Lambda\Sigma\frac{v^{2}_{\phi}r^{2}}{3}(\frac{1}{c^2})=
M\Sigma\frac{1}{r}-\Lambda\Sigma\frac{r^{2}}{3}
\end{array}
\end{equation}
Multiply equation (24) by $r$, then
\begin{equation}
\begin{array}{l}
\Sigma v^{2}_{\phi}r(\frac{1}{c^2})-2M\Sigma
v^{2}_{\phi}(\frac{1}{c^2})-\Lambda\Sigma\frac{v^{2}r^{3}}{3}(\frac{1}{c^2})
= MN-\Lambda\Sigma\frac{r^{3}}{3}
\end{array}
\end{equation}
Solving equations (24) and (25) for $\Lambda$ and $M$, we get
\begin{equation}
\begin{array}{l}
\Lambda=\frac{\Sigma
v^{2}_{\phi}(\frac{1}{c^2})-2M\Sigma\frac{v^{2}_{\phi}}{r}(\frac{1}{c^2})
-M\Sigma\frac{1}{r}}{\Sigma\frac{v^{2}_{\phi}r^{2}}{3}(\frac{1}{c^2})
-\Sigma\frac{r^{2}}{3}}
\end{array}
\end{equation}
And,
\begin{equation}
\begin{array}{l}
M=\frac{\Sigma\frac{v^{2}_{\phi}r^{3}}{3}(\frac{1}{c^2})\Sigma
v^{2}_{\phi}(\frac{1}{c^2})-\Sigma\frac{r^{3}}{3}v^{2}_{\phi}(\frac{1}{c^2})+\Sigma
v^{2}_{\phi}r(\frac{1}{c^2})\Sigma\frac{r^{2}}{3}-\Sigma
v^{2}_{\phi}r(\frac{1}{c^2})(\frac{1}{c^2})\Sigma\frac{v^{2}r^{2}}{3}}
 {2\Sigma
v^{2}_{\phi}(\frac{1}{c^2})\Sigma\frac{r^{2}}{3}-2\Sigma
v^{2}_{\phi}(\frac{1}{c^2})\Sigma\frac{v^{2}_{\phi}r^{2}}{3}(\frac{1}{c^2})-N\Sigma\frac{v^{2}_{\phi}r^{2}}{3}(\frac{1}{c^2})+
N\Sigma\frac{r^{2}}{3}+2\Sigma\frac{v^{2}_{\phi}r^{3}}{3}(\frac{1}{c^2})\Sigma\frac{v^{2}_{\phi}}{r}(\frac{1}{c^2})+
\Sigma\frac{v^{2}_{\phi}r^{3}}{3}(\frac{1}{c^2})\Sigma\frac{1}{r}}\\
\,\,\,\,\,\,\,\,\,\,\,\,\,\,\,\,\,\,\,\,\,\,\,\,\,\,\,\,\,\,\,\,\,\,\,\,\,\,\,\,\,\,\,\,\,\,\,\,\,\,\,\,\,\,\,\,\,\,\,\,\,\,\,\,\,\,\,\,\,\,\,\,\,\,\,\,\,
-2\Sigma\frac{r^{3}}{3}\Sigma\frac{v^{2}_{\phi}}{r}(\frac{1}{c^2})-
\Sigma\frac{r^{3}}{3}\Sigma\frac{1}{r}
\end{array}
\end{equation}
The expressions for $\Lambda$ and $M$ (equations 26 and 27) are
the functions of $r$ and $v$$_{\phi}$. Thus, the rotation curve
data ($v$$_{\phi}$, $r$) yield the value of the mass of the galaxy
($M$) and the cosmological constant ($\Lambda$). For the
computation, we use the software MATLAB6.1.
\begin{figure}
\vspace{0.0cm} \centering
\includegraphics[height=4.2cm]{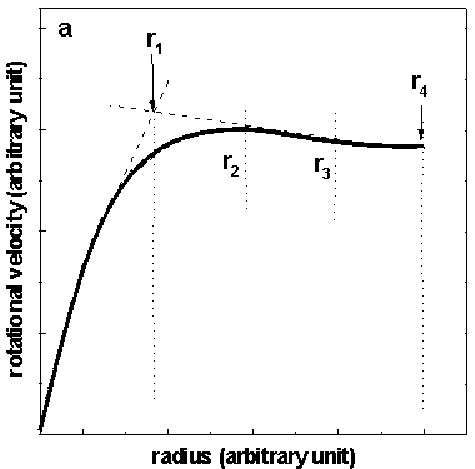}
\includegraphics[height=4.2cm]{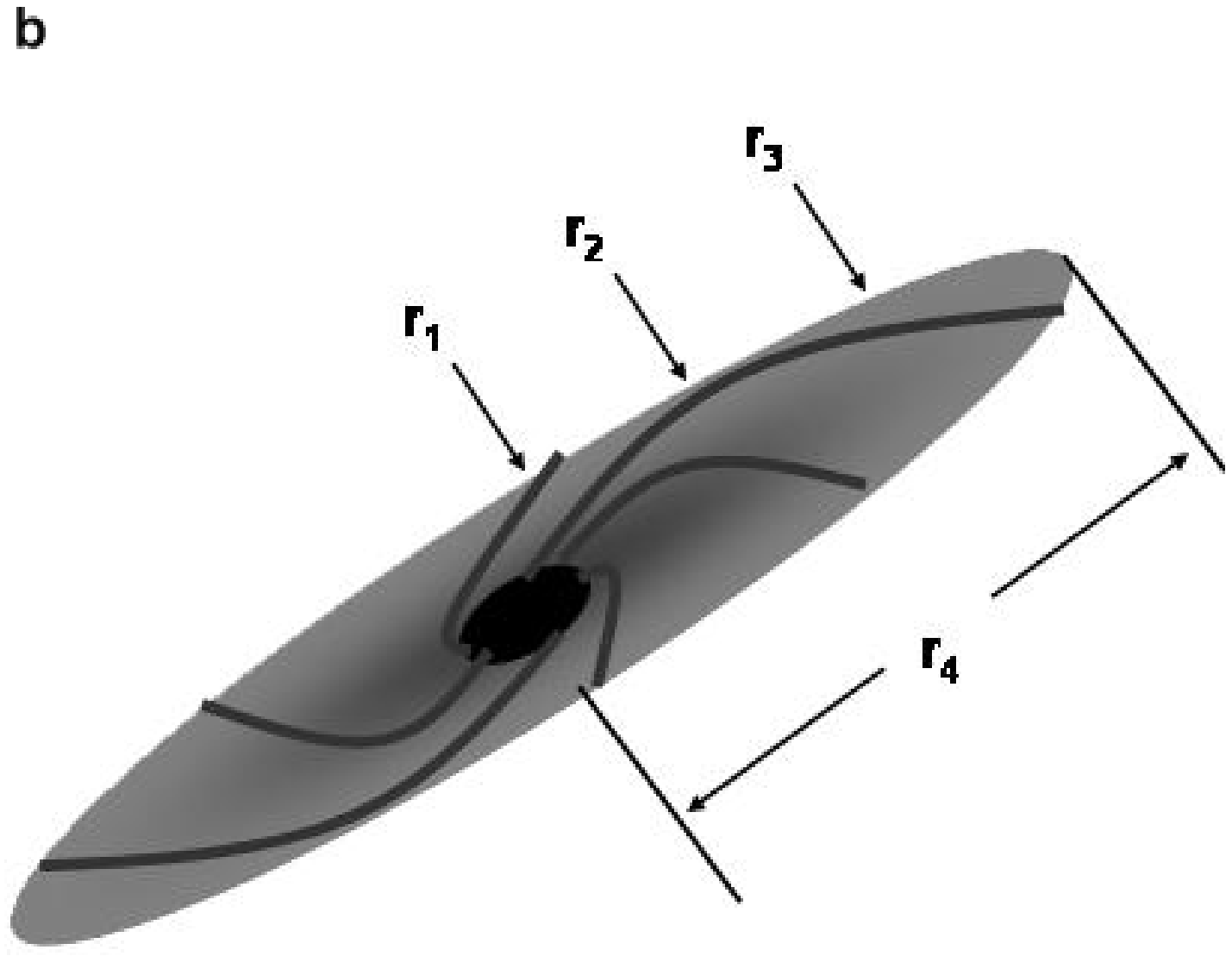}
\caption{(a) A typical sketch of the galaxy rotation curve. (b)
The radius $r_4$ represents the semi major diameter of the galaxy.
Other radii $r_1$, $r_2$, and $r_3$ are the different positions in
the flattened part of the galaxy rotation curve.}
\end{figure}
A typical sketch of the galaxy rotation curve is shown in Fig. 2.
In order to fix the positions (i.e., $r_{1}$, $r_{2}$, $r_{3}$ and
$r_{4}$) in the rotation curves, the flattened part of the
rotation curve is divided into 4 equal parts (see Fig. 2a). The
first position (say radius), i.e., $r_{1}$, is the point where the
flatness begins in the rotation curve. This point is fixed by
comparing the nature of all 15 galaxy rotation curves. This
position is found to be far from the central bulge of the galaxies
in our case. The position $r_{4}$ represents the extreme limit of
the radius from the center of the galaxy. We estimate the mass and
the $\Lambda$ of 15 spirals at the radii $r_{1}$, $r_{2}$, $r_{3}$
and $r_{4}$ (given in the third column of Table 2).

\section{Results}

Table 2 shows the values of the mass ($M$) and the cosmological
constant ($\Lambda$) of 15 sample galaxies at various radius
($r_{1}$, $r_{2}$, $r_{3}$, $r_{4}$). The mass of the galaxies are
found to lie in the range 0.13-7.60 $\times$ 10$^{40}$ kg (0.04 -
2.53 $\times$ 10$^{10}$ M$\odot$). These masses represent the
lower limit because the rotation curve data is not complete,
particularly of the central region and outer halo region. At the
central region, black hole can be expected. The dark matter is
believed to be dominating at the outer halo region. In the future
work, we intend to address this problem by modeling mass
distribution. The cosmological constant ($\Lambda$) is found to be
negative for all 15 sample galaxies.
\begin{table}
\caption{The first column lists the NGC name of the galaxy. The
next two columns represent the positions in the flat region of the
rotation curve. The fourth and fifth columns give the value of the
mass ($M$) and the cosmological constant ($\Lambda$).}
    $$
\begin{array}{p{0.08\linewidth}ccccc}
\hline \noalign{\smallskip}
NGC  & $positions$  &  r  &   M  &  -(\Lambda)   \\
            \noalign{\smallskip}
     &    & $(kpc)$ &   (\times 10^{40}\,\,$kg$) & (\times 10^{-40}\,\,$km$^{-2}) \\
            \hline
            \noalign{\smallskip}
0224  &   r_1/r_2/r_3/r_4  &   2.2/8.0/14.9/20.5   &   1.17/2.41/3.69/4.73    &   4.237/0.587/0.204/0.104   \\
0660  &   r_1/r_2/r_3/r_4  &   4.9/9.7/15.0/22.5   &   0.94/1.38/1.68/2.13    &   0.564/0.124/0.059/0.028   \\
0891  &   r_1/r_2/r_3/r_4  &   4.7/9.4/14.8/20.2   &   1.58/2.48/3.26/4.09    &   1.003/0.281/0.130/0.063   \\
1097  &   r_1/r_2/r_3/r_4  &   7.2/12.7/18.6/25.7  &   2.07/2.60/3.27/3.91    &   1.030/0.382/0.158/0.078   \\
1365  &   r_1/r_2/r_3/r_4  &   4.9/10.0/14.8/24.2  &   2.33/3.99/5.05/7.60    &   1.234/0.388/0.167/0.061   \\
2403  &   r_1/r_2/r_3/r_4  &   4.1/10.0/14.6/19.6  &   0.13/0.37/0.60/0.81    &   0.354/0.103/0.051/0.030   \\
2903  &   r_1/r_2/r_3/r_4  &   5.8/11.8/16.0/23.5  &   1.44/2.33/2.85/3.61    &   1.498/0.328/0.166/0.089   \\
3198  &   r_1/r_2/r_3/r_4  &   5.0/9.7/14.9/20.1   &   0.17/0.47/0.84/1.29    &   0.466/0.158/0.080/0.044   \\
4258  &   r_1/r_2/r_3/r_4  &   4.7/9.9/15.0/20.0   &   1.10/1.69/2.04/2.36    &   1.071/0.215/0.100/0.061   \\
4321  &   r_1/r_2/r_3/r_4  &   4.7/9.9/15.0/20.6   &   1.23/2.00/2.72/3.52    &   1.252/0.383/0.183/0.109   \\
4565  &   r_1/r_2/r_3/r_4  &   4.8/9.5/15.9/24.7   &   1.51/2.11/2.88/3.85    &   1.175/0.390/0.155/0.069   \\
5033  &   r_1/r_2/r_3/r_4  &   4.0/8.9/14.7/22.2   &   1.27/1.95/2.53/3.24    &   2.219/0.458/0.194/0.096   \\
5055  &   r_1/r_2/r_3/r_4  &   4.7/9.8/14.8/21.4   &   0.55/0.96/1.35/1.78    &   1.159/0.298/0.123/0.056   \\
5236  &   r_1/r_2/r_3/r_4  &   4.7/9.8/14.8/21.2   &   0.88/1.20/1.53/1.90    &   0.760/0.218/0.093/0.046  \\
5907  &   r_1/r_2/r_3/r_4  &   3.5/9.5/14.8/20.7   &   0.76/1.42/2.28/3.14    &   1.795/0.402/0.181/0.106   \\
\noalign{\smallskip} \hline
\end{array}
     $$
\end{table}
A comparison between the variation of mass and the radii (i.e.,
positions away from the bulge of the galaxies) is shown in Fig.
4a. As expected, it is found that the mass of the galaxy increases
with the increase of the radius of the galaxy. This increase is
found to be steeper for massive galaxies (Fig. 4a). This is
probably due to the huge size of the halo of the massive galaxies.
This suggests that the mass of the halo depends upon the size of
the galaxies. A good agreement can be seen for blue-shifted galaxy
(dashed line in Fig. 3a).

Fig. 3b shows the radius versus $\Lambda$ plot. As the positions
($r_{1}$, $r_{2}$, $r_{3}$, $r_{4}$) from the center of the galaxy
increase the $\Lambda$ is found to decrease (in magnitude). In
other words, one can expect positive $\Lambda$ at greater $r$
(probably, outside the galactic halo). In Fig. 3b, the width of
the distribution in the lower radius side (i.e, $r_{1}$) is wider
than that of the higher radius (i.e., $r_{4}$) side. Probably, the
positive $\Lambda$ demands the convergence of these distributions.
We noticed that the variation of $\Lambda$ is minimum when moving
out of the bulge of the galaxy.

It can be concluded that the $\Lambda$ within galactic halo
corresponds to the dark energy which tries to de-accelerate the
accelerating Universe. In addition, one can predict that the
$\Lambda$ outside the halo might be positive, supporting supernova
cosmology project. A similar result is noticed for the
blue-shifted galaxy (dashed line in Fig. 3b)
\begin{figure*}
\vspace{0.0cm} \centering
\includegraphics[height=4.0cm]{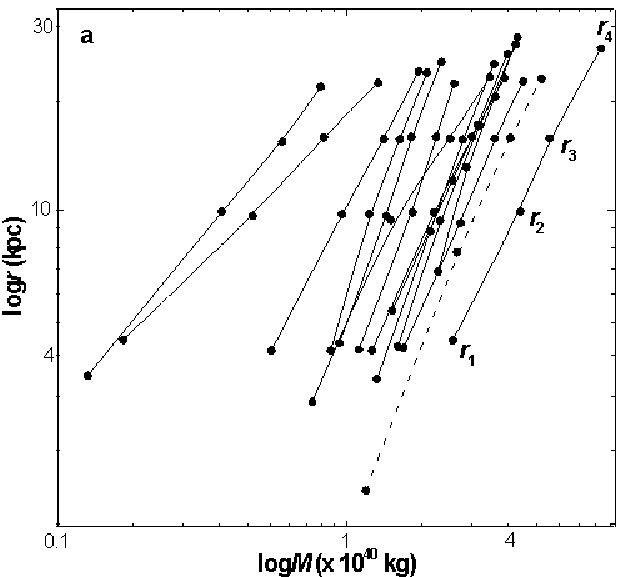}
\includegraphics[height=4.0cm]{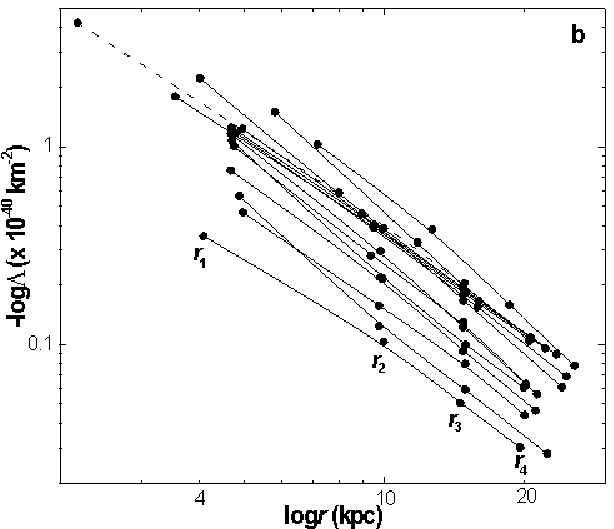}
\includegraphics[height=4.0cm]{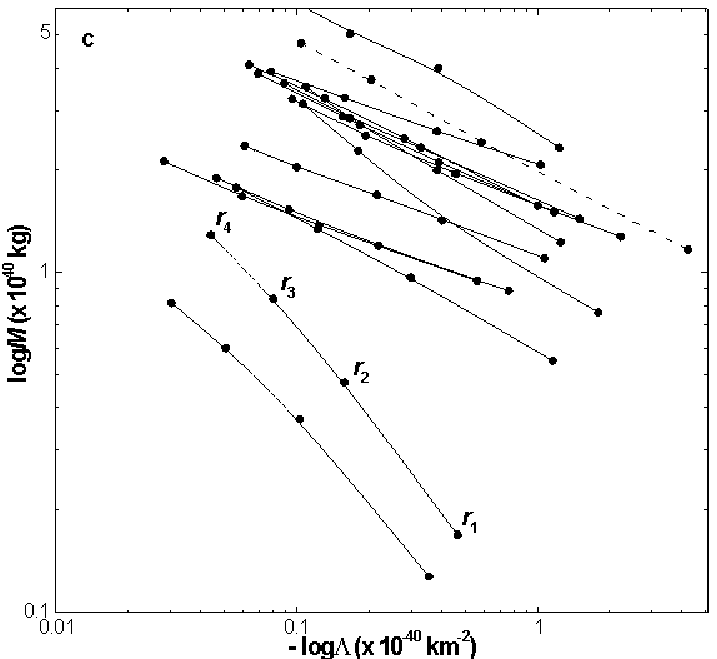}
\caption{The variation between the radius ($r$) (a,b), mass ($M$)
(a,c) and the cosmological constant ($\mid\Lambda\mid$) (b,c) for
15 sample galaxies. The solid circles represent the estimated
value. The observed values are fitted polynomially (second order).
The solid line is for the red-shifted galaxies and the dashed line
is for the blue-shifted galaxy NGC0224.}
\end{figure*}
The $\Lambda$ versus $M$ plot for all 15 galaxies can be seen in
Fig. 3c. As $r$ increases, as expected, mass of the galactic halo
($M$) increases. Probably, this increase lead the $\Lambda$ tend
to go to positive. However, we found negative $\Lambda$ for our
sample galaxies. Our galaxies are nearby galaxies. It would be
interesting to analyze the flattened portion of the rotation
curves of high red-shift galaxies. We intend to work on rotation
curves of $\sim$100 galaxies that have RV $>$ 5\,000 km $^{-1}$ in
the future. We expect to find positive $\Lambda$ for distant
galaxies.

\section{Discussion}

Kraniotis \& Whitehouse (2000) estimated $\Lambda$ value for 6
spiral galaxies using Friedman energy equation (Sciama 1995;
Bahcall et al. 1999), weak field approximation (Whitehouse \&
Kraniotis 1999), and the observed rotation curve of the galaxy.
They found that the rotation curve data of all 6 galaxies up 50
kpc (see Table 1 of their paper) inevitably lead to a negative
value of the cosmological constant, suggesting de-accelerating
Universe.

We have included all 6 galaxies (NGC2406, NGC4258, NGC5033,
NGC5055, NGC2903, NGC3198) analyzed by Kraniotis \& Whitehouse
(2000) in our database in order to study the difference between
the results. In our study, the $\Lambda$ value for these galaxies
at $\sim$ 20 kpc are found in the range $-$0.03 to $-$0.09
$\times$ 10$^{-40}$ km$^{-2}$. They estimated $-$0.5 to $-$5.5
$\times$ 10$^{-55}$ cm$^{-2}$. Their range for $\Lambda$ is
relatively wider than ours.

Our results as well as results obtained by Kraniotis \& Whitehouse
(2000) contradict the result obtained from the Supernova Cosmology
project. It has to be kept in mind that the Supernova project
studied the recessional velocity of distant objects while in this
paper, we are concerned with the rotational velocity of galaxies.
Obviously, the bound matter in galaxies can not be accelerating
with the expansion of the universe. So during the formation of
galaxies local effect must have worked in such away that, an
effect of negative vacuum energy was generated in the region.
Probably, the vacuum energy of these structures lead to form a
nearby local bubble. The space-time of this bubble might be anti
de Sitter type. We suspect that the dark energy adds to the
attractive gravity of matter within the bubble in order to show
de-accelerating feature. We are focusing our future works on this
very aspect.

\section{Conclusion}

We used an expression for rotational velocity of a test particle
in a circular motion around the central mass in an asymptotically
de Sitter space time (Pokheral 1994) and estimated the value of
the $\Lambda$ from the rotation curve data of 15 spiral galaxies.
We conclude the following:

\begin{enumerate}

\item The mass of the galaxies are found to lie in the range
0.13-7.60 $\times$ 10$^{40}$ kg (0.04 - 2.53 $\times$ 10$^{10}$
M$\odot$).

\item The cosmological constant ($\Lambda$) is found to be
negative ($-$0.03 to $-$0.10 $\times$ 10$^{-40}$ km$^{-2}$) for
all 15 galaxies. These are nearby (radial velocity less than
1\,636 km$^{-1}$) galaxies. A similar result is noticed for a
blue-shifted galaxy, NGC0224.

\item As the radius of the galaxy increases, mass of the galactic
halo increases and the $\Lambda$ tend to increase from larger
negative value to the smaller values. The $\Lambda$ within
galactic halo corresponds to the dark energy which cause to
de-accelerate the accelerating Universe.

\item Our results show that the relative effect of cosmological
constant ($\Lambda$) becomes pronounced at larger distances, away
from the bulge of the galaxy.

\end{enumerate}

The result of this work do not agree with the results of supernova
project which looked at the large scale cosmology. We suspect that
the $\Lambda$ value might be positive outside the local bubble,
supporting supernova cosmology project. The space time of this
local bubble might be anti de Sitter type. Such anti de Sitter
space-time is becoming important in the context of the brane
world. We intend to justify this prediction in our future works.


\section{Acknowledgments}
This research has made use of the NASA/IPAC Extragalactic Database
(NED) which is operated by the Jet Propulsion Laboratory,
California Institute of Technology, under contract with the
National Aeronautics and Space Administration. We are grateful to
Prof. R. Weinberger, Innsbruck University for his constructive
criticism.


\end{document}